\documentstyle[prl,aps,epsf,twocolumn,palatino,palatcm]{revtex}
\begin{document}
\tighten
\draft
\narrowtext
{\bf Comment on  ``Shell Filling and Spin Effects in a Few Electron Quantum
Dot''}\\

The experimental observation of shell and spin effects in the
ground-state
addition
spectrum of a semiconductor quantum dot has been reported in a recent
Letter by Tarucha {\em et al.\/} \cite{tarucha96}.
By applying a gate voltage, the
chemical potential of the dot is varied and one observes Coulomb
oscillations in the current vs gate voltage. 
Current peaks indicate the tunneling of the $N+1$st
electron into the $N$-electron dot 
when the chemical potential $\mu_N=E(N+1)-E(N)$ 
gets into resonance with the source
and drain potentials. 

The authors show the addition
spectrum vs magnetic field for $N=0$ to 23. They explain the
observed structures by a model of non-interacting electrons in a
disc with parabolic lateral confinement $V(r)={1\over
2}m^\star\omega_0^2r^2$ ($m^\star=0.067\,m_0$ in GaAs). 
A confinement energy
of $\hbar\omega_0\approx 3$ meV fits well to the measured minimum of the
7th-electron addition energy
at $B\approx 1.3$ T, which they attribute to
the crossing of the 6th and 7th single-particle energy levels.
Spin effects are interpreted in the framework of  
a constant-interaction model with
fitting parameters including the exchange energy of parallel spins.

Using current- and spin-density functional theory (CSDFT) in
local-density
approximation (LDA) \cite{vignale,ferconi}, 
we have calculated the
addition-energy spectrum of a parabolic circular quantum dot in a
perpendicular
magnetic field with up to 20 electrons.
Our calculations give a satisfactory
agreement with the experiment (see Fig.~1).
This is remarkable, since
we use a mean-field approach 
and calculate the exchange-energy by employing the
local spin-density approximation. Thus no further 
fitting
parameters appear besides\ $\hbar\omega_0$ for the bare confinement
(in contrast with the model in Ref.~\cite{tarucha96}).    
The latter turns out to be larger (5 meV) than what the experimentalists
assumed for their fit. The reason for this difference is that the bare
confinement is screened by the interacting electrons.
It is also interesting to regard our calculations 
as a test of the LDA for a small number of electrons
in the dot.

Due to interaction, the shell structure of the non-interacting
model (connected with ``magic numbers'' $N=2$, 6, 12 etc.) 
at zero magnetic field 
is partially broken. In the Kohn--Sham (KS) scheme of CSDFT, 
only the single-particle angular-momentum states with
opposite
signs are still exactly degenerate. Nevertheless, we obtain a spectrum
with increased addition energies for the 7th and 13th electron as seen
in experiment at $B=0$.

We confirm the observed Hund's rule for small magnetic fields which
favors spin alignment. In Fig.~1 this leads to a reduction of the
addition energy if the electron is placed into the same spin
state as its predecessor
(e.g. for $N=4, 8$ or 14) and vice versa ($N=5, 9$ or 15). 

Quasi-degenerate spin states at intermediate magnetic fields give rise
to pairs of parallel addition-energy traces (indicating spin
saturation for even $N$) \cite{tarucha96}.
Deviations from this behavior occur when
the spin-saturated system is near a level
crossing of different angular-momentum states at the Fermi energy. 
Then again, Hund's rule applies since it becomes energetically
favorable to occupy different
angular-momentum states and align spins
in order to gain exchange energy. As a result, the addition of, 
e.g., the 7th electron around $B=1.2$ T [Fig.~1(a)] is especially
costly where the exchange gap of the 6-electron dot is the largest
[cf. Fig.~1(b)]. Correspondingly,
the addition-energies
for the 5th and 6th electron (now spin-aligned) are no longer parallel.
This effect has not
been discussed in Ref.~\onlinecite{tarucha96} although it is (weakly)
visible in Fig.~2 of Ref.~\onlinecite{tarucha96}
for $N=6\rightarrow 7$ at $B\approx 1.3$ T
and occasionally more obvious for larger $N$.

\unitlength1cm
\begin{picture}(5,13)
\put(-0.3,0){\epsfysize13.3cm\epsfbox{adden_tot.epsi}}
\end{picture}
\begin{figure}
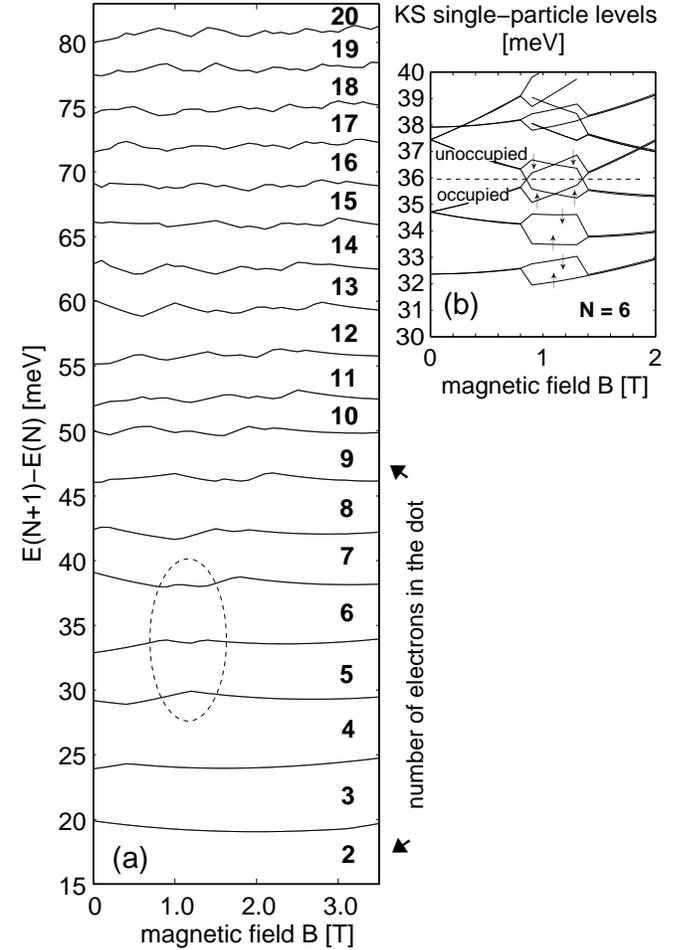

\caption{(a) Addition spectrum vs magnetic field for a circular 
dot with 
$\hbar\omega_0=5$ meV and $N=2...20$. This
is to be compared with Fig.~2 in Ref. \protect\cite{tarucha96}.
The encircled region is discussed in the text.
(b) KS energy levels for $N=6$ showing the exchange
gap around $B=1.2$ T. Arrows indicate spin states.
}
\end{figure}

{\small\noindent
Oliver Steffens and Ulrich R\"ossler\\
\hspace*{3mm}
Institut f\"ur Theoretische Physik, Universit\"at Regensburg,
\hspace*{3mm}
D--93040 Regensburg, Germany
\vspace{5mm}\\
\today\\
PACS number(s): 73.20.Dx, 71.15.Mb, 73.23.-b

\end{document}